# A Method for Vehicle Collision Risk Assessment through Inferring Driver's Braking Actions in Near-Crash Situations

Liqun Peng, Miguel Angel Sotelo, IEEE Fellow, Yi He, Yunfei Ai, Zhixiong Li, IEEE Member

*Abstract*—**Driving information and data under potential vehicle crashes create opportunities for extensive real-world observations of driver behaviors and relevant factors that significantly influence the driving safety in emergency scenarios. Furthermore, the availability of such data also enhances the collision avoidance systems (CASs) by evaluating driver's actions in near-crash scenarios and providing timely warnings. These applications motivate the need for heuristic tools capable of inferring relationship among driving risk, driver/vehicle characteristics, and road environment. In this paper, we acquired amount of real-world driving data and built a comprehensive dataset, which contains multiple "driver-vehicle-road" attributes. The proposed method works in two steps. In the first step, a variable precision rough set (VPRS) based classification technique is applied to draw a reduced core subset from field driving dataset, which presents the essential attributes set most relevant to driving safety assessment. In the second step, we design a decision strategy by introducing mutual information entropy to quantify the significance of each attribute, then a representative index through accumulation of weighted "driver-vehicle-road" factors is calculated to reflect the driving risk for actual situation. The performance of the proposed method is demonstrated in an offline analysis of the driving data collected in field trials, where the aim is to infer the emergency braking actions in next short term. The results indicate that our proposed model is a good alternative for providing improved warnings in real-time because of its high prediction accuracy and stability.**

## I. INTRODUCTION

D riving safety is a high priority issue for governmental agencies, the majority of vehicle manufacturers and other stakeholders. In order to enhance the safety for both the drivers and pedestrian, a number of improvements have been proceeded, ranging from enhancement of infrastructure to vehicle-based safety systems. Recent advanced driver assistance systems (ADAS) techniques have made a large-scale sensor embedded vehicle study to collect amounts of driving data on the actual road to investigate the relationship between the emergency driving safety and driver maneuvers (e.g., Acceleration/ deceleration, and steering) [1]. This approach was described as most helpful in studying driver behavior and accident-causation-mechanism. With access to field driving data, the safety-related events could be observed and measured more precisely.

*Problem motivation:* Effective ADAS requires awareness of actual driving situation, a reliable assessment of the vehicle crash risks, and making rapid decisions on assisting actions [2][3]. On the one hand, understanding of the multi-factors on road, especially the driver behavior, will remarkably improve the vehicle crash risk assessment. For example, if the speed of the vehicle under study is 90 km/h and the relative distance from the vehicle ahead is 50 m, the acceleration volition would be considered as dangerous/risky, conversely, if the driving volition is slow-down, the risk level is low and therefore the action should be considered as not dangerous. On the other hand, although a variety of roadside and vehicular on-borad sensors are capable of collecting a large-scale information, it is still needed to be considered whether all these collected data are appropriate for traffic safety applications. The inclusion of abundant factors for crash detection may lead to overfitting actual driving safety and making false warnings to drivers.

*Approaches for vehicle collision risk assessment:* Driving involves complex interactions between the driver, the vehicle, and the environment under varying conditions (road characteristics and properties, driver intention, incidental effects, etc.). However, the correlation of real-time driver behavior with the driving safety in a short time period has hardly been demonstrated in previous studies, especially when considering the entire complexity of situations in the context of driving. Furthermore, the less siginifcant characteristics of driver behavior, vehicle or traffic environment take negative efforts for driving safety estimation and prediction. From this perspective, it appears that existing methods integrating multiple factors to judge safety is not satisfied enough to realistically model real-world safety issues.

In this work, we investigate the actual driving behaviors in near-crash events as well as the involved interactions between the driver, vehicle and traffic environment under risky and safety conditions. The near-crash events were identified in field experiments by detecting unusual vehicle kinematics using accelerometers and gyroscopic sensors equipped in the test vehicle, described in Section "III. Data collection and processing". The driving data was collected under potential threats in dynamic traffic, and a comprehensive dataset is built to record all related "driver-vehicle-road" factors as a whole. In the present paper, we propose a reasoning model to infer the vehicle collision risk involved in an emergency event. First, the vehicle crash risk is graded by the braking process characteristics, namely high deceleration, medium deceleration, and low deceleration respectively. Then, an improved variable precision rough set (VPRS) model is applied to define fixed rules, which profile the near-crashes into different risk levels based on the three aforementioned braking process features. Then, information entropy model is employed to evaluate the significance of multiple "driver-vehicle-road" attributes to derive a representative index reflecting the driving risk for each situation. By detecting the upcoming dangerous driving events, we are able to provide driver with immediate feedback assisting driver to adapt safety maneuver.

The main contribution of this paper, which is the novel method proposed, will be fully described in Section "IV. Modeling process". One of the goals of our completed study is to critically analyze the input and output factors that have been taken into consideration in the analysis of driving situation and risk, and to determine which are the most critical factors that ADAS should address in their next generation. We extend current situation and risk prediction approaches with other factors, such as, for instance, consideration of human individual characteristics. Another goal of our research is to develop a driver-adaptive reasoning model, to prototype it as a built-in driving assistance system, and to test it in practical situation.

The reminder of this paper is organized as follows. Section II presents a brief overview of the related research on vehicle collision avoidance system. Section III introduces field driving test and data collection, including experiment design, data processing and driving risk definition. Section IV describes the modeling process for vehicle crash risk assessment. The model evaluation and test results are illustrated in Section V. Finally, the main conclusions and the future work are discussed in Section VI.

## II. RELATED WORK

A variety of driving safety assessment have been explored. Some have evaluated the safety issue mainly based on real-time vehicle kinematics. Others have comprehensively tried to monitor the D-V-E (driver, vehicle, and environment) statuses. In this study, situation and risk evaluation methods is the focus, which accounting for more complex scenarios and factors.

### A. Crash risk assessment based on vehicle kinematics

Safety distance (SD) model is one of the most important methods in identification of longitudinal crash risk [2]. As presented in Fig. 1, where $v_1$ and $v_2$ are the longitudinal velocity of the following and preceding vehicles respectively, $d_r$ is the gap between them, it is intended to calculate the critical warning distance $d_w$, which could be expressed as the general function form as follows:

$$d_w = f(v_{rel}, v_1, \alpha_1, \alpha_2, \tau) + d_0 \qquad (1)$$

Where $v_{rel}$ is the relative velocity between the following and preceding vehicles, $\tau$ is the delay, and $\alpha_1$ and $\alpha_2$ are the maximum deceleration of the following and leading vehicles respectively, $d_0$ is headway offset, the variables represented above are comprehensively taken into consideration when evaluating the safety distance $d_w$. Then, by comparing the measured headway $d_r$ with the safety distance $d_w$, the vehicle will be in the safe driving situation when $d_r > d_w$, while the following vehicle should be warned or decelerated to avoid a crash if $d_r < d_w$. The safety distance model could be transformed into time to collision (TTC) model [3-5]. Both SD and TTC models have been extensively applied in many modern developed in-vehicle safety systems based on Information and Communication Technology [6][7]. Such systems have been expected to support the driver to maintain safe speed and headway in all driving situations by providing timely warning to driver when a critical safety situation emerges.

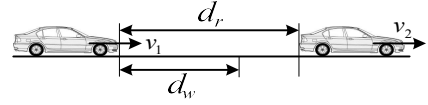

Fig. 1. Safety distance analysis scheme

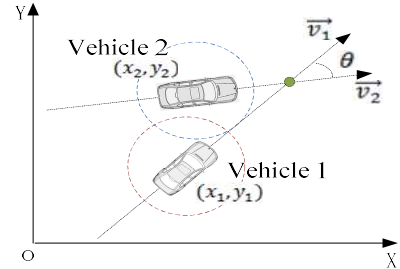

Fig. 2. Two dimensions analysis scheme

Other researchers developed two dimensions algorithms based on dynamic equations of vehicle motion. Considering the vehicles points in space, as present in Fig. 2, implies that collision risk exist when

$$\sqrt{(x_1(t_1) - x_2(t_2))^2 + (y_1(t_1) - y_2(t_2))^2} \leq \sum_{i=1}^{2} R_i \qquad (2)$$

where

$$R_i = g(x_i, y_i, \overline{v_i}, L_i, W_i)$$

$L_i$ and $W_i$ is respectively the length and width of the subject vehicle and the obstacle. The safety zones algorithm creates a safety virtual zone around vehicles and detects the overlap areas between the subject vehicle and each approaching obstacle to indicate collision danger [8][9]. However, the algorithms based on vehicle kinematics are very susceptible to generate false warnings, especially when driver behavior is ignored in analyzing complex traffic scenarios.

However, aforementioned studies were typically done by setting absolute thresholds on the vehicle kinematics measurements, without taking account of the relationship between the crash risk severity and detailed driving maneuver and (e.g., constant speed, acceleration, braking, and steering).

### B. Crash risk assessment based on D-V-E arrangement

It is well known that driving involves complex interactions between the driver, the vehicle and the environment under varying conditions (road characteristics and properties, weather, incidental effects, etc). The D-V-E (driver, vehicle, traffic environment) factors have been generally considered to be the most important factors in crash occurrence [10][11]. Hence, It is necessary to develop reasoning models, as shown in Fig. 3, which integrate the main constituents of driving situation with generic phases of completing driving, i.e. perception, analysis, decision making and action. Such a reasoning model will improve the current development of driving safety analysis.

Naturalistic driving studies provide an opportunity to more precisely observe and measure safety-related events [12-14]. In these studies, the driver's factor was fully considered as one of the precipitating and contributing factors of crashes and provided the critical exposure of pre-crash data. Naturalistic driving studies have recorded a large-scale field data, which in turn, could provide a useful supplement to effectively control laboratory and field studies to further enhance the understanding of the effects of driver characteristics on traffic safety [15-17].

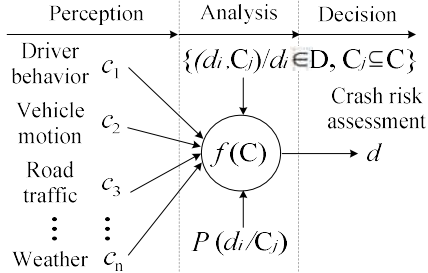

Fig. 3. Crash risk assessment with consideration of multi-factors

Other studies presented a wider survey of the D-V-E arrangement, taking into consideration driving safety related factors, such as obstacle detection, driver intention, real-time weather and roadway geometry [18-20]. For instance, Hassan et al., [21] used the structural equation modeling approach to explore significant factors associated with young drivers' involvement in at-fault crashes. It was revealed in the study that, aggressive violations, in-vehicle distractions and demographic characteristics were significant factors affecting young drivers' involvement in at-fault crashes. Ahmed et al., [22] also assessed the effectiveness of the weather on real-time road crash risk in locations with recurrent fog problems. Although some studies have made effort to address these issues by combining multiple elements (e.g., detecting the driving context, analysis of conditions, and proposing actions), their number and genuine contribution were relatively low.

### C. Crash risk assessment inference reasoning model

Machine learning methods have been widely studied to identify crash risk using relevant vehicle motion and dynamic traffic. [23] illustrated that using multi-roles as inputs of machine learning predictor significantly reduced the number of false collision warnings to drivers compared to the analytically derived formula based on the minimum safety gap. It also explicates the advantage of machine learning models capable of training a large volumes of collected data to form a concise and meaningful understanding of actual driving situation. Forming a high-level view of the world is a necessary requirement for intelligent vehicles to interact safely with both human drivers as well as other intelligent vehicles [24].

Artificial neural network model is one of the most practical tools used for fitting the relationships between risk driving behavior and traffic environmental factors, which model vehicle collision risk as a complicated nonlinear function of "driver-vehicle-road" attributes. The nonlinear function is defined by a multilayer network, including one or two hidden layers, with "driver-vehicle-road" attributes as inputs and collision risk prediction as output [25]. It is believed that a three layer neural network with a sufficiently large number of hidden neurons can model any nonlinear relationships between inputs and outputs [26]. There is variety of evolving neural network algorithms illustrated for improving application, [27] processed with a large number of inputs from accelerometer and gyro measurements based on a self-organized neural network model. The proposed approach is capable of recognizing dangerous conditions though heuristically tuning thresholds from simulated training crash tests, which outperforms the benchmark method by setting absolute thresholds on the inertial measurements. [28] proposed a probabilistic neural network is trained to predict prospective steering angles based on collected video data and the vehicles CAN bus data during human driving, thus imitating human behavior. The integration of end-to-end learning into a modularized architecture allows for additional safety constraints and complementary sensor information to be combined with intuitive steering.

Logit-based model is another popular methodology for analyzing crash the factors associated with accident severity. For example, [29] developed an unsupervised and Bayesian model that generates local multivariate linear models describing how the risky driver behavior is associated with the input data (independent variables), which are segmented into blocks of linear data sequences based on local statistical patterns. The advantage of this model formulation is using matrix-variate distribution theory, providing a general, intuitive and flexible parameterization. [30] employed a tree-based rules to analyze accidents involving powered two-wheelers, and demonstrated that the curve alignment, rural areas, run-off-the-road crashes, nighttime, and rainy weather were significantly associated with accident severity. These studies provided some insights into the factors that affect the likelihood of a vehicle crash. [31] obtained new insights into driving risk by using classification and regression tree (CART) model to analyze the relationship of driver characteristics, road conditions, and vehicle characteristics in near-crash database. The results indicate that the velocity when braking, triggering factors, potential object type, and potential crash type had the greatest influence on the driving-risk level involved in near-crashes. It also evaluated the application of CART model for predicting motor vehicle crashes, and showed that CART model performed better than traditional decision tree models.

In process of machine learning algorithms, the data may include easily hundreds of variables, a key question therefore whether or not all these variables actually lead to true information gain? The answer is obviously, no, since there are a lot of redundant variables that may increase the performance of the learning dataset but they do not necessarily increase the performance on the actual validation dataset which can be easily controlled for by keeping an eye on the over-fitting. Many machine learning techniques such as neural networks, tree-based models, and support vector machines perform worse when extra irrelevant predictors are added, and therefore a variable selection technique should always precede the modeling [32]. Rough set based models are some of the most practical tools, which are highly resistant to the inclusion of irrelevant variables through automatic variable subset selection. The use of rough set theory is widely advocated for the sake of building a very effective classifier for real world traffic event detections. It is mainly used for the discovery of data dependencies, the evaluation of the importance of attributes, the discovery of data patterns, and the reduction of all redundant objects and attributes to a minimal representative subset of attributes [33-35]. Attributes reduction is very important in rough set based data analysis because it can be used to simplify the induced decision rules without reducing the classification accuracy [36][37]. One of the main advantages of rough set based models is their simple interpretability. In spite of Artificial neural network models having strong nonlinear fitting capabilities, their input-output relationships cannot be interpreted or verified explicitly, whereas the rough set based fitting model explicitly indicate the input-output relationships characterized by reduced rules, which are interpretable and easy to understand [38]. The transparent input-output relationships are very important to retro designing collision warning strategy, especially evaluate the significant factors impacting the driving safety in emergency cases.

## III. DATA COLLECTION AND PROCESSING

In this section, a specified field test was carefully designed and performed to collect real driving data using naturalistic and low-intervention methods, which was used to analyze the driving safety in near-crash scenarios under complex road traffic environment.

### A. Experiment design (Test vehicle/sensors/drivers/route)

The field driving experiments were conducted using a YUEXIANG sedan, which was provided by CHANGAN Auto company. The vehicle was equipped with instruments to detect driver behavior, vehicle motion state, and dynamic traffic in real time situation. The on-board units equipped in the experimental vehicle includes two cameras, Mobileye, two radars and on-board computer, as shown in Fig. 4. The two cameras were used to record vehicle forward view and driver's facial expression. Mobileye was used to identify the road traffic environment (lane lines and obstacles information preceding vehicle) and judge vehicle crash risk by detecting TTC (time to collision). The two radars were used to measure the headway between vehicle and approaching vehicles in front and behind respectively. On-board computer was used to record data obtained by sensors, including GPS, brake signal, steering signal, three-axis acceleration information.

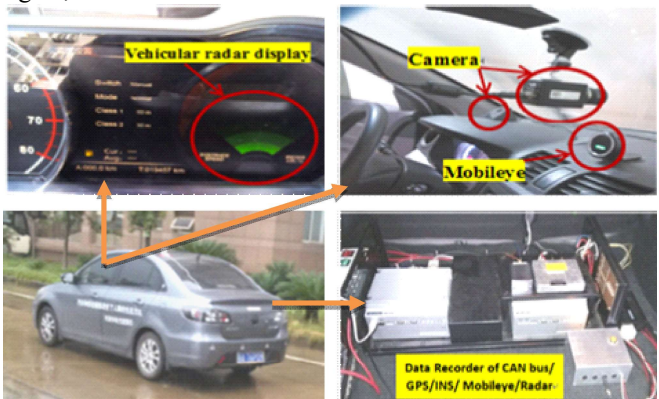

Fig. 4. Experimental vehicle and apparatus

Field trip was completed by our experiment vehicle equipped with above mentioned on-board units. Testing route was designed surrounding over the area in the central part of the Wuhan city, China, as shown in Fig. 5. The Google Map image records the test route (solid line) where the data was collected by vehicle with on-board GPS equipment. However, the GPS raw data described in longitudinal and lateral degree could not be directly used for evaluating the vehicles' trajectory in the travelled distance. More so, the corresponding RTK (real time kinematic) positions in Fig. 6 represents the vehicle trajectory in plane-coordinate. The points of origin and destination have been remarked. The RTK positions represent the vehicle trajectory in the test area.

The selected route for driving test is representative of most urban city traffic conditions in China, i.e., city ring road and expressway (usually low traffic volume and may have congestion). The experiments were implemented from 7:30 am to 9:30 am and 17:00 pm to 19:00 pm. Within these time frames, the traffic flow is denser and traffic crash is more frequent. In this study, the driver's high deceleration behavior was considered to be a crash risk related event. A totally of 51 drivers, who signed the consent form, participated in the designed driving experiments. The experiment lasted 60 days on average 4 hours per day, during which, the driving time and range was approximately 265.8 hours and over 5101.79 km respectively. Among the 51 drivers, 6 were female and 45 were male, all the participants held a valid driving license. The average age was 37 years (ranging from 25 to 56). They had 12 years (ranging from 3 to 16) mean period of driving experience.

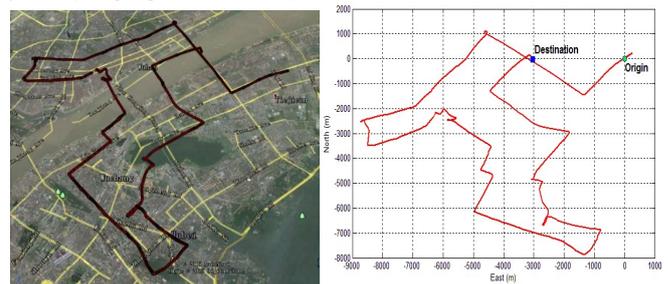

Fig. 5. The route of the area of one on-road test

### B. Dataset Processing

This research focuses on the driving safety analysis and assessment in near crash scenarios. Driving risk is identified as a potential threat that could cause vehicle collision accidents. Usually, the consequence of driving risk for a driver in his/her normal state is mainly reflected by rapid evasive maneuvers (i.e. emergency braking and/or steering operation), which have been employed by many studies on naturalistic driving to identify near-crashes situations [10-12]. Near crash implies that the driver performs a rapid evasive maneuver (i.e. emergency braking and /or steering operation) that did not result in real crash. In the experiments, near crash events in naturalistic driving were identified by detecting unusual vehicle kinematic. When the vehicle deceleration reached a threshold value (longitudinal -1.5 m/$s^2$, lateral: -1 m/$s^2$) or TTC (time to collision) between the test vehicle and preceding vehicle is less than 3s, the data collection system recorded the vehicle state (i.e., speed, brake signal, steering signal, and three-axis acceleration), the TTC with approaching vehicles in the longitudinal direction, and video sequence of the events happening at the time. Note that, it is very necessary to review the recorded video data to decide whether an event triggered by kinematic thresholds was actually safety critical. If not, such an event was not defined as near-crash and was deleted from the dataset. The recorded cases were checked manually.

Totally, 3374 near-crash events (only in the longitudinal direction, with 1 real crash accident) were recorded throughout the 30 days real driving test. Nearly all the near-crashes had large longitudinal deceleration, implying that the drivers tended to adopt the rapid braking maneuver to avoid potential crash. Hence, the driving-risk level was represented by the braking process characteristics. Intuitively, the driving risk is higher if the braking maneuver is performed with greater urgency in a near-crash. The clustering braking process characteristics data were investigated to evaluate the involvement of driving risk in a near-crash event [20]. The distribution of these near-crashes by deceleration is summarized in Table 1. The driving-risk level in each near-crash case will be placed in one of the following three groups: low-risk, moderate-risk and high risk.

TABLE 1 DISTRIBUTED CATEGORY OF NEAR-CRASH RISK

| Driving risk level | Low | Moderate | High |
|---|---|---|---|
| Deceleration when braking m/s$^2$ | (-2, 0] | (-5, -2] | (-8, -5] |

As outlined in previous studies, driver behavior, vehicle motion and traffic environment have been largely investigated and testified as among the major factors influencing driving safety in varying degrees. In this study, we conduct a reasoning model to predict driver's response and action in near-crash situation. It incorporates procedures that (1) detect the driving environment and to extract safety related conditional information about it, the status of the vehicle, and the conditions of the driver. It usually performs the sub-processes for monitoring, detection and classification of the information, for recognition of driver behavior and environmental factors and vehicle's actual states (position, orientation, conditions), (2) analyze the driving situation and conditions, which is to achieve the comprehension of the driving safety and to project it to the whole of the driving arrangement and process. Mathematically, the reasoning model can be defined as: IF $C_1^k \wedge C_2^k \wedge \ldots \wedge C_T^k$, THEN $\{(D_1, \beta_{k1}), (D_2, \beta_{k2}),\ldots, (D_N, \beta_{kN})\}$ with $\{\beta_{kj} \leq 1 | j = 1,2,\ldots,N\}$, a rule weight $\theta_k$ and attributes weight $\{\omega_i | i = 1,2,\ldots,T\}$, where $\{(D_1, \beta_{k1}), (D_2, \beta_{k2}),\ldots, (D_N, \beta_{kN})\}$ is referred to as a reliability $\beta_{kj}$ of the inferred results for each $D_j$. The belief rule can be understood as if $C_1=C_1^k$, $C_2=C_2^k$, $\ldots$, $C_T=C_T^k$, then the consequence is THEN $\{(D_1, \beta_{k1}), (D_2, \beta_{k2}),\ldots, (D_N, \beta_{kN})\}$, where $C_1$, $C_2$,$\ldots$, $C_T$ are conditional attributes of the inference rule, $D_1, D_2,\ldots, D_N$ are assessment grade used in the consequence, and $\beta_{kj}$ is the belief degree to which $\{D_j | j = 1,2,\ldots,N\}$ is to be believed to be the consequence. (3) predict the threats and consequences of that threat based on the comprehensive similarity of current conditional attributes $C_1^{k+1} \wedge C_2^{k+1} \wedge \ldots \wedge C_T^{k+1}$ with attributes in each inference rule. The prediction can be expressed as $D_{k+1} = \{D_j, \max_{N \geq j \geq 1} \beta_{kj} | \beta_{kj} = \theta_k * \sum_{i=1}^{T} \omega_i S(C_i^k, C_i^{k+1})\}$. Based on the weighted similarity results for each attribute, the prediction of the sample $C_1^{k+1} \wedge C_2^{k+1} \wedge \ldots \wedge C_T^{k+1}$ can be assessed using decision in $D_j$ that maximizes the $\beta_{kj}$. From the above, we can apply the inference model in particular for assessing driving safety status with comprehensive consideration to driver behavior, vehicle motion and road environment. Then the prediction result is responsible for generating warnings for the driver and for the execution of the corrective actions, depending on the level of risk.

Altogether, the experiment dataset included the following five major categories: Participants information (age, gender, driving experience); Driver behavior and decision (acceleration, deceleration, steering); Road obstacles (time to collision in longitudinal direction); Vehicle kinematic status (velocity); Road traffic (traffic flow, road segment, road slipperiness). The dataset presented above is comprehensive and contains important attributes that describe the conditions affecting vehicle crash risk. It also provides potential information for analyzing the relationship among driving risk, driver/vehicle characteristics, and road environment. However, each attribute in the dataset has been defined and described by a specific performance measure (i.e. vehicle velocity is expressed by km/h, vehicular approaching status is expressed by Time to collision (s). Driver age is expressed by years and driver braking action is expressed by Boolean signal (1 or 0)). These could not be used directly for comprehensive evaluation by integrating other items with different property unit. In order to unify the property of the above attributes, a quantitation protocol is proposed in Table 2, with explicitly considering the factors distribution in Chinese traffic situations and the distribution statistics of crash accidents [11]. Then, the heterogeneity among the attributes presented above can be eliminated, which in turn, can be applied for comprehensive analysis. It should be noted that, the quantitation range of attribute quantification need to be covered according to the real application.

TABLE 2 QUANTITATION OF ATTRIBUTES

| Attributes | Type | Description |
|---|---|---|
| **Participants information** | | |
| Gender | Boolean | 1: Male; 2: Female |
| Age | Continuous | Driver age (years) is categorized into four groups, 1: 18-30; 2: 31-45; 3: 46-60; 4: >60 |
| **Driver behavior** | | |
| Acc pedal | Boolean | 0: No; 1: Yes |
| Brake switch | Boolean | 0: No; 1: Yes |
| Turn indicator | Boolean | 0: No; 1: Yes |
| **Road obstacles** | | |
| Vehicular distance with approached obstacle in longitudinal direction | Continuous | Evaluated by TTC (time to collision, seconds) and quantified into three levels: 1: >5; 2: 2.1-5; 3: 0-2 |
| **Vehicle kinematic status** | | |
| Velocity | Continuous | Evaluated by km/h, quantified into four levels: 1: 0-40; 2: 41-50; 3: 51-60; 4: >60 |
| **Road Traffic** | | |
| Road segment type | Qualitative | 1: Corridor link; 2: Intersection; 3: Viaduct; 4: Tunnel |
| Traffic flow | Qualitative | 1: Congested; 2: Moderate flow; 3: Free flow |
| Road slipperiness | Continuous | Evaluated by coefficient of friction between tyre and road surface, quantified into three levels: 1: 0.7-1; 2: 0.4-0.69; 3: 0-0.39 |

Note for Driver behavior Boolean descriptions: Further categorized into: 1: Keep constant; 2: Acceleration; 3: Deceleration; 4: Steering

## IV. MODELING PROCESS

In this section, we proposed an improved rough set model for: (1) evaluating the collected attributes that influence the driving safety in real traffic environment, (2) predicting the vehicle collision risk in near-crash scenarios. Fuzzy sets and rough sets are widely applied for modeling vagueness and uncertainty issues, such as driver behavior. However, fuzzy set reasons whether driver behavior possibly belongs to the set or rarely related mainly based on subjectively assigning membership function value [35], while rough set processes the problem by investigating the sample data and estimating the conditional probabilities related to a special many-valued logic. In this case, no additional operators are pre-defined and classical set-theoretic operators are referred to define rough set operators[36].

To achieve the accurate prediction, a hybrid VPRS (variable precision rough set) and Information entropy model is investigated explicitly for mining the field test data and categorizing the near-crash driving situation in corresponding safety level. The correlation of driving safety with all types of attributes is revealed and analyzed, and significance of relevant attributes, such as driver decision, vehicle motion and traffic environment, on the influence of driving risk is evaluated.

### A. RS and VPRS for data mining

Rough Set (RS) is an effective approach for addressing problems of data classification, based on the conception of upper and lower approximation in a Decision Table (DT), which are constructed from empirical data and can represent the correlation of condition factors with decision factors [36]. The DT is characterized by four tuple set $S = \{U, A = C \cup D, V, f\}$, where $U = \{x_1, x_2, ..., x_{|U|}\}$ denotes a non-empty finite set called universe, $A$ is a nonempty finite set of attributes that contains condition attribute set $C = \{a_1, a_2 ..., a_m\}$ and decision attribute set $D = \{d_1, d_2, ..., d_n\}$. $V = \cup V_a$ is the value domain of the attribute $a$, which represent the properties of either condition attributes or decision attributes. $f: U \times A \to V$ is a total function such that $f(x_i, a_j) \in V_a$ for every $\forall x_i \in U$ and $\forall a_j \in A$, e.g., $f(x_i, a_j) = v$, which means for element $x_i$, its attribute $a_j$ has the value of $v$. For an arbitrary nonempty subset $B \subseteq A$, an indiscernibility relation is defined as:

$$IND(B) = \{\{x_i, x_j\} \in U * U / f(x_i, a) = f(x_j, a), \forall a \in B\}$$

$IND(B)$ partial $U$ into a family of disjoint subsets $U/IND(B)$ called a quotient set of $U$:

$$U/IND(B) = \{[x]_B / x \in U\}$$

where $[x]_B$ denotes equivalence class determined by $B$. Then for a decision table (DT), the indiscernibility class of $U$ with regards to condition attribute set $C$ can be expressed as $[x]_C = \{c_1, c_2, ..., c_m\}$, and with regards to decision attribute set D can be expressed as $[x]_D = \{d_1, d_2, ..., d_n\}$. The relationship between condition attribute set $C$ with decision attribute $d_j$ can be evaluated by lower approximation and upper approximation, which are defined as:

$$\begin{cases} \underline{apr}_C(d_j) = \cup \{x \in U | [x]_C \subseteq [x]_{d_j}\} \\ \overline{apr}_C(d_j) = \cup \{x \in U | [x]_C \cap [x]_{d_j} \neq \emptyset\} \end{cases} \quad (3)$$

The positive region of $d_j$ to $C$ is defined as $POS_C(d_j) = \underline{apr}_C(d_j)$.

Variable precision rough set (VPRS) is proposed as an important extension of classical RS. The VPRS gives a less rigorous definition of the inclusion relation compared with Eq. (1), which makes the classical RS more fault tolerant. By introducing a precision parameter value $\beta \in (0.5, 1]$, the lower and upper approximations of $d_j$ can be defined as follows:

$$\begin{cases} \underline{apr}_C^\beta(d_j) = \cup \{x \in U | P(d_j | C) \geq \beta\} \\ \overline{apr}_C^\beta(d_j) = \cup \{x \in U | P(d_j | C) > 1 - \beta\} \end{cases} \quad (4)$$

where $P(d_j | C)$ is the inclusion degree function:

$$P(d_j | C) = \frac{|[x]_C \cap [x]_{d_j}|}{|[x]_C|} \quad (5)$$

The improved positive region of $d_j$ to $C$ is defined as

$$POS_C^\beta(d_j) = apr_C^\beta(d_j). \tag{6}$$

The classification quality of VPRS is evaluated by classification quality degree. If the decision attribute set divides the $U$ into $n$ classes, the classification quality degree of a certain attribute set $P$ ($P \subseteq C$) can be defined as follow:

$$\gamma^\beta(P,D) = \frac{\sum_{i=1}^n |apr_P^\beta(d_j)|}{|U|} \tag{7}$$

$\gamma^\beta(P,D)$ presents the percentage of effective sorting decision information $D$ based on $P$ in certain knowledge set, e.g., $\gamma^\beta(P,D) = 0$, it denotes that condition attribute set $P$ includes no significant factors related to decision attributes in $D$. Noted that $0 \leq \gamma^\beta(P,D) \leq 1$.

### B. β-reducts for VPRS attribute reduction

For certain DT, not all of the condition attributes included is effective for information system category, which means that, some of the condition attributes are redundant. The condition attributes reduction in DT is one of the core problems for both VPRS. The process of finding the reduct is to identify the important attributes and remove the redundant attributes from condition attribute set in a certain DT. Formally in VPRS, $\beta$-reducts of condition attributes is expressed as $red^\beta(C,D)$, which should be satisfied with the following two properties:

1. $\gamma^\beta(C,D) = \gamma^\beta(red^\beta(C,D),D)$,
2. No proper subset of $red^\beta(C,D)$, subject to the same $\beta$ value can also give the same quality of classification.

The parameter $\beta$ can be interpreted as confidence value, on which, the largest proportion of condition equivalence classes $[x]_C$ can be allocated correctly to different decision equivalence classes $[x]_D$. $\beta$-reducts derived subsets of the attributes, which are capable, through construction of decision rules, of explaining allocations given by the whole set of condition attributes subject to the majority inclusion aspect. The $\beta$-reducts should ensure the reduct-derived decision rules compatible with those from the original DT. Two propositions should be clearly considered to investigate the allowable ranges of $\beta$ for attributes reduction, which allow the subsets of condition attributes to remain $\beta$-reducts [37].

**Proposition 1** If condition attribute set $C$ is discernible to $d_j$ with a certain $\beta$ value between $(0.5, 1]$, then the $C$ is also discernible at any level between $(0.5, \beta)$.

**Proposition 2** If condition attribute set $C$ is not discernible to $d_j$ with a certain $\beta$ value between $(0.5, 1]$, then the $C$ is also not discernible at any level between $(\beta, 1]$.

Thus, the confidence level associated with a set of attributes is defined by the least upper bound value on $\beta$ such that all the condition classes satisfy the majority inclusion relation at this value. The least of these upper bounds of the $\beta$ is defined as:

$$\beta = min(m_1, m_2) \tag{8}$$

where

$$\begin{cases} m_1 = 1 - max\{Pr(d_j|c_i) \,|\, P(d_j|c_i) < 0.5\} \\ m_2 = min\{Pr(d_j|c_i) \,|\, P(d_j|c_i) > 0.5\} \end{cases}$$

The definition operates in terms of the quality of classification, which is used to define and extract $\beta$-reducts. The requirement is that a $\beta$-reduct should permit the use of subsets of attributes without loss of classification quality. $\beta$-reducts extract significant attributes from decision table and build rules for classification of unseen samples by matching the description of the sample to the condition part of each rule.

### C. Attributes weighted similarity for decision making

Decision making rules extracted by VPRS reduct cannot cover the complete cases, where the sample does not match any of the rules. That is, if the matched rule is certain, it is clear that the class of the sample can be evaluated using the decision of the matched rule. However, if the matched rule has not been included, the classification is ambiguous. In this section, we propose weighted attributes to evaluate extracted rules, and design a decision algorithm based on attributes weights similarity to classify an unseen sample.

Suppose $U/X = \{X_1, X_2, \ldots, X_l\}$ is an equivalence class produced by a set of condition attributes $X$, $X \subseteq C$. $U/D = \{Y_1, Y_2, \ldots, Y_{|D|}\}$ is an equivalence class produced by decision attribute set $D$. The information entropy of subset $X$ can be defined as [36][37]:

$$H(X) = -\sum_{i=1}^l \frac{|X_i|}{|U|} log_2(\frac{|X_i|}{|U|}) \tag{9}$$

The conditional entropy of $D$ given $X$ is defined as:

$$H(D/X) = -\sum_{i=1}^l \frac{|X_i|}{|U|} \sum_{j=1}^{|D|} \frac{|Y_j \cap X_i|}{|X_i|} log2(\frac{|Y_j \cap X_i|}{|X_i|}) \tag{10}$$

The mutual information entropy of $D$ to $X$ is defined as:

$$I(X,D) = H(D) - H(D/X) \tag{11}$$

If $X_i \in X$, the significance of $X_i$ to the classification results can be evaluated by:

$$SIG(X_i, X, D) = abs(\triangle I) \tag{12}$$

Where, the $abs(\triangle I)$ is the absolute value of the $\triangle I$, which is the mutual information degree. It is defined as:

$$\triangle I = I(X,D) - I(X - \{X_i\}, D) = H(D/X - \{X_i\}) - H(D/X) \tag{13}$$

If $X_p, X_q \in X$, the relative significance of $X_p$ to $X_q$ can be evaluated as:

$$SIG_{p,q} = SIG(X_p, X, D)/SIG(X_q, X, D) \tag{14}$$

Suppose $B = \{b_1, b_2, \ldots, b_{|B|}\}$ is condition attributes set in extracted rules, the corresponding weights set for each condition attribute is $\omega_b = \{\omega_{b_1}, \omega_{b_2}, \ldots, \omega_{b_{|B|}}\}$, which is calculated by the geometric average method as follows:

$$\omega_{b_i} = (\prod_{q=1}^{|B|} SIG_{i,q})^{1/|B|} \tag{15}$$

After normalizing, the weight of each condition attribute is presented as follows:

$$\varepsilon_i = \frac{\omega_{b_i}}{\omega_B}, \quad \omega_B = \sum_{i=1}^{|B|} \omega_{b_i} \tag{16}$$

If the sample does not match any of the rules, the decision algorithm based on attributes weighted similarity as shown below could be used to deal with the remaining cases.

Suppose that $u_i \in B, u_j \in U$, the similarity of the $u_i$ and $u_j$ on attribute $b_i$ is defined as:

$$S_{b_i}(u_i, u_j) = 1 - \frac{|v_i - v_j|}{|b_{max} - b_{min}|} \tag{17}$$

In Eq. (14), $v_i$ and $v_j$ denotes the value of attribute $b_i$ in $u_i$ and $u_j$ respectively. $b_{max}$ and $b_{min}$ respectively denotes the maximum and minimum of attribute $b_i$ in $B$. Using the weighted similarity measurement to evaluate the similarity of $u_i$ and $u_j$ as follows:

$$S(u_i, u_j) = \frac{1}{n} \sum_{i=1}^n \omega_i S_{b_i}(u_i, u_j) \tag{18}$$

In Eq. (18), $n$ is the number of rules in $B$. Based on the similarity result, the class of the sample $u_j$ can be assessed using decision in $u_i$ that maximizes $S(u_i, u_j)$.

From the above, the improved rough set model examines the driving safety with comprehensively evaluating the state of driver behavior, vehicle motion and road traffic, and extracting rule sets for matching the sample conditions to each classification of driving safety. Then, the significance of each factor will be evaluated by mutual information entropy and the weights are calculated based on the unseen sample in the real field situation that would be finally classified into decision by matching the similarity of the sample to the condition part of each rule.

## V. EVALUATION AND DISCUSSION

### A. Illustrative examples

#### 1) Extreme Driving Event Detection

In this section, the details of driver behavior and vehicle motion are visualized. Then, extreme braking events will be identified in accordance with special rules. One of the rules this study used is TTC-based thresholds [2], since driving behaviors vary at different TTC situation, implying different driving safety contexts. Further, these extreme braking events were linked to instantaneous vehicle control statuses and driving contexts to understand why they occur. Note that this study uses driver's extreme acceleration as key measures to capture driver's instantaneous collision avoiding decisions, i.e., how a vehicle is maneuvered instantaneously. Driving behaviors can also be captured by other measures, such as steering angles and the position of the accelerator or brake in a vehicle. Given that accelerations are the outcomes of maneuvering by drivers, the authors prefer to use them for analysis.

In our work, we extracted 678 groups of samples from field test, as described in Section 3.1, and take vehicle longitudinal emergency cases as example to explicitly evaluate the crash risk of the test vehicle and preceding vehicle in near-crash scenarios. This subset is a representative sample, in which the experimental vehicle recorded all the parameters in Table 2 when the vehicle deceleration reached a threshold of -1.5 $m/s^2$ or the TTC less than 3s, the immediate data and previous sampling points were both recorded. Then, an offline behavior analysis is performed by randomly divided these samples into two subsets: 628 groups of these data were used for searching a β-reduct decision table (DT) of condition attributes which provide the same information for classification purposes as the full set of available attributes, then the significance of potential risk factors on driving safety can be evaluated and quantified based on the β-reduct DT by taking advantage of mutual information entropy for assigning attributes weights. The other 50 groups of data were extracted for inferring the extreme brake maneuvers happened in next 0.5s by integrating the weighted β-reduct attributes as inputs, the results of predicted braking extent, actual driver deceleration and real time headway (TTC) for this case are summarized in Fig. 6.

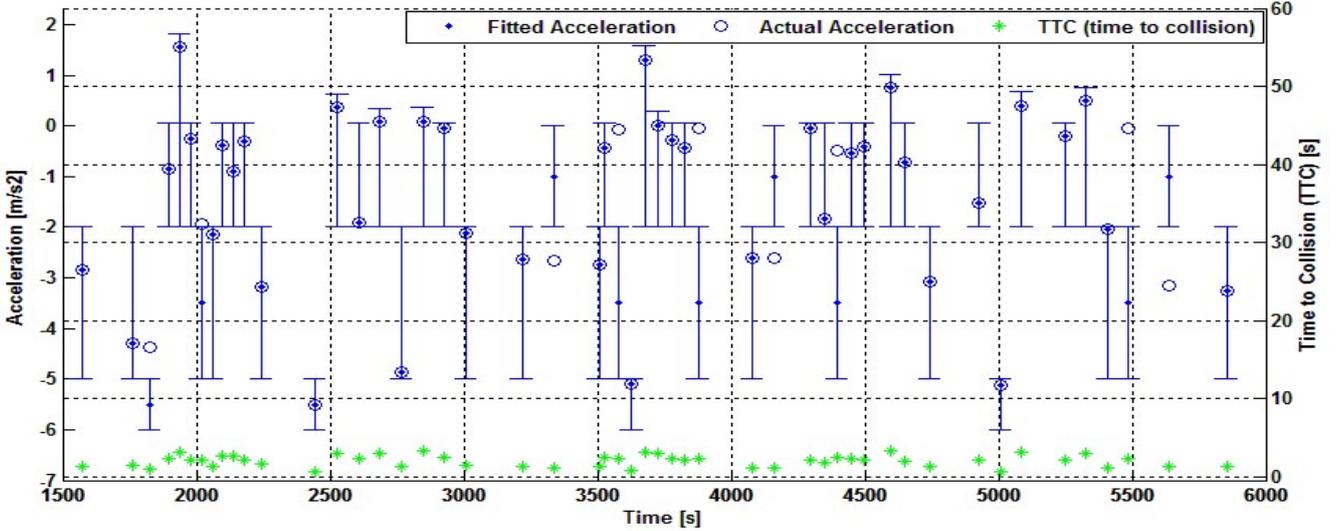

Fig. 6. Crash risk evaluation of longitudinal near-crash scenarios

#### 2) General Results

In Fig. 6, the actual driver acceleration/deceleration are represented with circle point and classified into three scopes (m/s²): $(-2, 1.8]$, $(-5, -2]$ and $[-6, -5]$, which respectively indicate three crash risk levels. The inferred driver's acceleration/deceleration in next short term are represented with solid dot. The longitudinal headway between vehicle on changing lane and approaching vehicles in neighbor lane is also evaluated by TTC, which is usually widely accepted as binomial judgement for assessing vehicle crash risk by setting a threshold [2][3]. For example, it was identified that, at t = 1571s, 2768s, 3628s, etc., the TTC between the test vehicle and preceding vehicle was less than 2s, these scenarios are viewed as risk situation.

Our proposed method examines driver intension, vehicle motion state as well as the approaching vehicles in dynamic traffic, and infer the driver's compulsory reaction for safety driving according to the current vehicle crash risk. Fig. 6 shows a comparison of our predicted results and actual driver deceleration, the results confirm that the combination of "driver-vehicle-road" based classifier produces more accurate predictions and few errors outperform the classifier using only TTC. It is observed that, at t = 1827s, the TTC between the test vehicle and preceding vehicle was less than 0.6s, and the driver was predicted to apply a hard breaking, which represent high risk situation. Actually, that was the only real collision accident during the whole test period. However, at t = 3218s, the longitudinal headway (TTC) is 1.2s, which indicate the short distance to the preceding vehicle, and this near-crash is correctly identified as moderate risk (the consequent deceleration was -2.64 m/$s^2$). In this case, the driver was detected taking a timely deceleration action in advance. Consequently, the driver could only take a moderate deceleration to avoid collision accident. It was also noticed that, at t = 3628s, where the TTC is 0.92s, but we predict that the high risk will come to the next short term (the actual deceleration was -5.09 m/$s^2$), since the driver took an acceleration action before recognizing the potential crash risk and taking a harsh brake to avoid the accident, which was evaluated to be a high risk situation. It illustrates the influence of driver behavior and decision on the driving safety.

*B. Scoring comparison*

This section explains how the model performance are evaluated. We analyze the driving risk of near-crash scenarios in the naturalistic driving experiments. Near-crash implies that the driver performs a rapid evasive maneuver (i.e., emergency braking and/or steering operation), failing which a real crash may occur. Previous studies indicated that these evasive driving events are associated with comprehensive "driver-vehicle-road" conditions, the main component of vehicle crash assessment models are interpreting the factors importance and understanding their relationship with driving safety. The attributes subset $C = \{c_1, c_2, c_3, c_4, c_5, c_6, c_7, c_8, c_9\}$ extracted from field trial have been investigated for this study, where the $c_1$ denotes driver behavior or intention, $c_2$ denotes the gender of test driver, $c_3$ denotes the age of test driver, $c_4$ denotes vehicle velocity, $c_5$ denotes TTC in occupied lane, $c_6$ denotes TTC in neighbor lane, $c_7$ denotes road segment type, $c_8$ denotes traffic congestion, $c_9$ denotes road slipperiness. However, in process of machine learning algorithms, the data may include easily hundreds of variables, a key question therefore whether or not all these variables actually lead to true information gain? The answer is apparently not, since redundant variables may increase the performance of the learning dataset but they do not necessarily increase the performance on the actual validation dataset which can be easily controlled for by keeping an eye on the over-fitting. A scoring process is conducted to examine the accuracy and reliability of our proposed method for extreme driving events detection, it is also compared with the results derived by PNN algorithm, CART algorithm and TTC algorithm respectively.

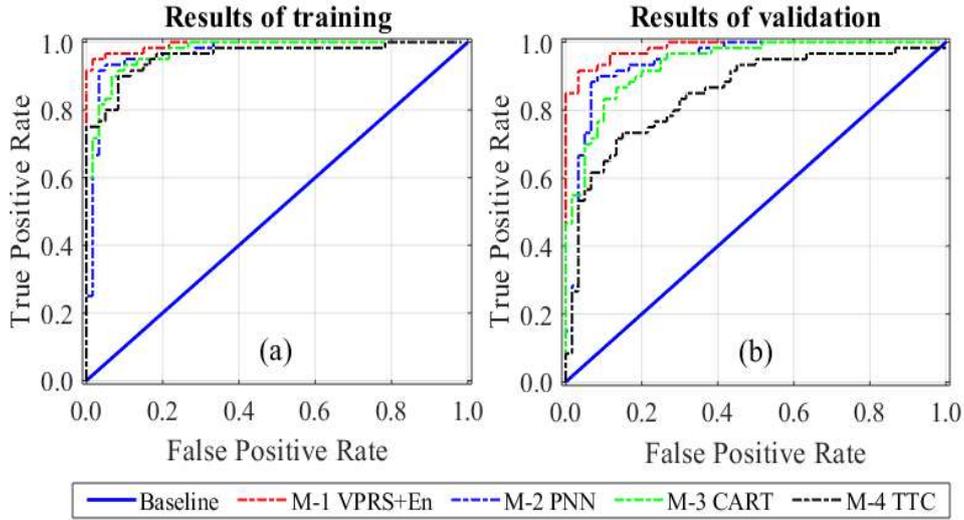

Fig. 7 Overall performance of crash risk assessment models

Table 3 Validation: classification rates and index

| Algorithms | Description of Inputs | TPR (%) | FPR (%) | TNR (%) | OCR (%) | AUC |
|---|---|---|---|---|---|---|
| M-1: VPRS+En | Reduct attributes | 91.7 | 3.3 | 96.7 | 94.2 | 0.94 |
| M-2: PNN | All collected attributes | 88.3 | 6.7 | 93.3 | 90.8 | 0.88 |
| M-3: CART | Normalized attributes | 83.3 | 10.0 | 90.0 | 86.7 | 0.84 |
| M-4: TTC | Vehicular TTC | 71.7 | 13.3 | 86.7 | 79.2 | 0.71 |

*1) Model Accuracy Evaluation*

The scoring comparison were conducted for four different model. Model-1 was calibrated using rough set reduct attributes vector $\{c_1, c_4, c_5, c_6, c_9\}$ as input according the results of our proposed model. In order to examine the prediction accuracy that can be achieved depending only on one dataset at a time and to account for significance of reduct element from the collected data source, another three models were calibrated and compared; Model-2 based on PNN algorithm using all available factors gathered from field trial as model inputs; Model-3 based on CART algorithm consider experiment gathered factors as inputs, of which, the relative importance of all variables are normalized to characterize their ability to influence the model; Model-4 based on considering the threshold of TTC $\{a_5\}$ as criteria for determining the risk level.

The Receiver Operating Characteristics (ROC) curve is capable of examining the classification problem with positive and negative class values [37]. Through plotting a True Positive Rate (TPR) vs False Positive Rate (FPR) graph, it shows how well the model is at discriminating between the crash and non-crash cases in the target variable. In our study, the low vehicle crash risk is defined as the negative class and the high & moderate crash risk is defined as the positive class. We use ROC curve indexes as the main criteria to examine the performance of models for vehicle crash risk detection. The TPR is the ability to predict a crash case correctly and True Negative Rate (TNR = 1-FPR) is the ability to predict a non-crash case correctly. The overall accuracy indicates the proportion of correctly identified positive and negative cases, and the area under the ROC curve (AUC) represents the expected performance as a single scalar. The exact criteria for all models validation datasets are listed in Table 3.

Consequently, Model-1 is consistently superior in term of classification accuracy and area under the ROC curve. Model-2 is ranked second after the full model (Model-1), while Model-3 is relatively ranked lower than Model-1 and Model-2 but still providing satisfactory performance. Model-4 is ranked the lowest on these measures. Area under the ROC curves as shown in Fig. 7 and listed in Table 3 was found to be 0.94 for Model-1 validation dataset, 0.88 and 0.84 for Model-2 and Model-3, respectively while Model-4 achieved ROC of 0.71 all for the validation dataset. It may also be observed that Model-1 achieves 91.7% correct prediction of driver harsh deceleration in consequent short term by using the β-reduct attributes as input, while only 71.7% of vehicle crash risk has been predicted by using Model-4 (TTC model). It indicates the significance of driver volition $a_4$ and weather condition $a_9$ on the impact of safety driving. Although the attribute $a_4$ and $a_9$ have no direct relativity with vehicle crash risk, when comprehensively consider all attributes, the prediction has been improved, which testify the over speeding behavior and road snippiness effectively characterize the potential vehicle crash risk. We further conduct the prediction based on Model-2 and Model-3 respectively, and examine the prediction performance involved all the attributes $C$ as inputs, then we achieve the less accurate results of 88.3% and 83.3% driver deceleration maneuver when compared to the performance of using β-reduct attributes, which account for the redundant attributes $a_2$, $a_3$ and $a_8$ having insignificant impact on driving safety.

It should be noted that the overall accuracy and error rate are particularly suspicious performance measures when the class distribution of a data set strongly biases to the majority class. Highly imbalanced problems generally have highly non-uniform error costs that often favor the minority class of primary interest. For instance, identifying a dangerous driver behavior as safe may be a fatal error, while identifying a safe driving behavior as risk is usually considered a much less serious error since this mistake can be corrected in later detections. ROC graphs are consistent for a given problem even if the distribution of positive and negative samples is highly skewed. The comparisons of model predictions between the observed and predicted risk levels for the learning and testing data are also presented in Fig. 7. The overall Model-1 prediction accuracy for the learning data is approximately 95.9% and that for the testing data is approximately 94.2%, which is a more optimal range compared with the other training models. For instance, in Fig. 7, we investigate the PNN model for vehicle crash prediction and achieved accuracies of 93.9% and 90.8% in the training and testing phases, and achieved accuracies of 89.6% and 86.7% for training and testing procedure based on CART model. The prediction performance of our proposed model demonstrates that the VPRS model structure can reflect the pattern hidden behind naturalistic data to some extent.

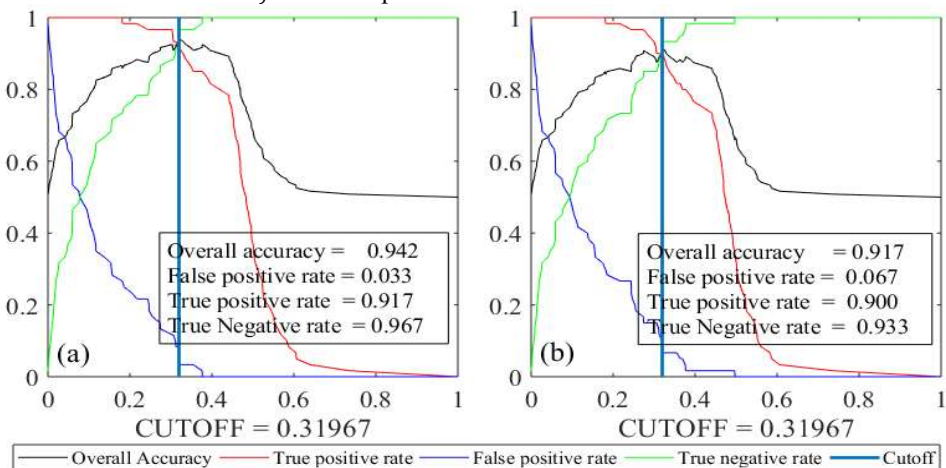

Fig. 8. Model-1 vehicle crash risk pre-detection

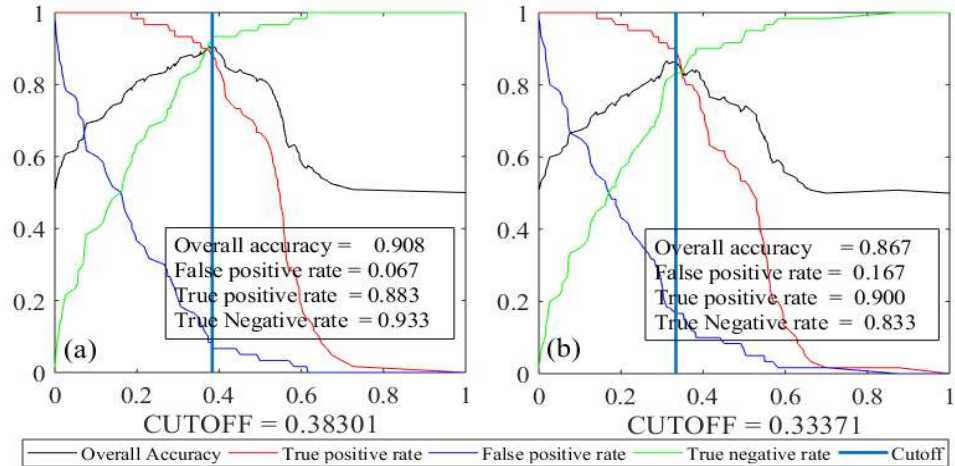

Fig. 9. Model-2 vehicle crash risk pre-detection

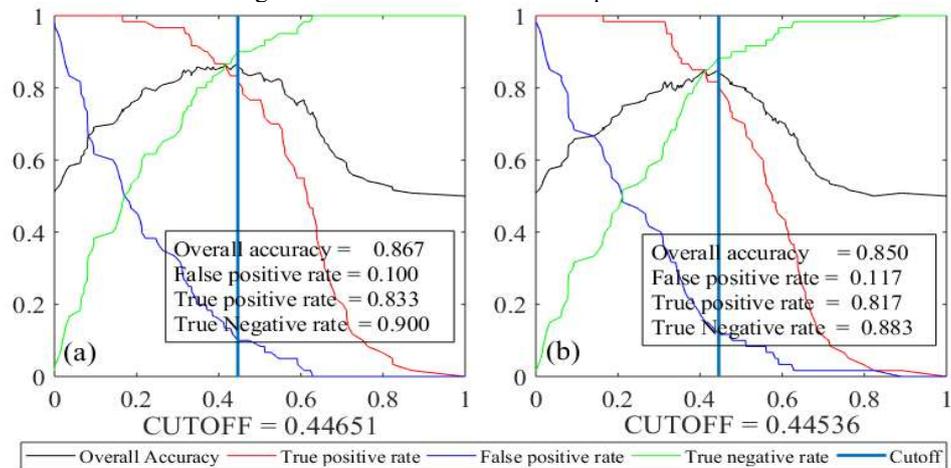

Fig. 10. Model-3 vehicle crash risk pre-detection

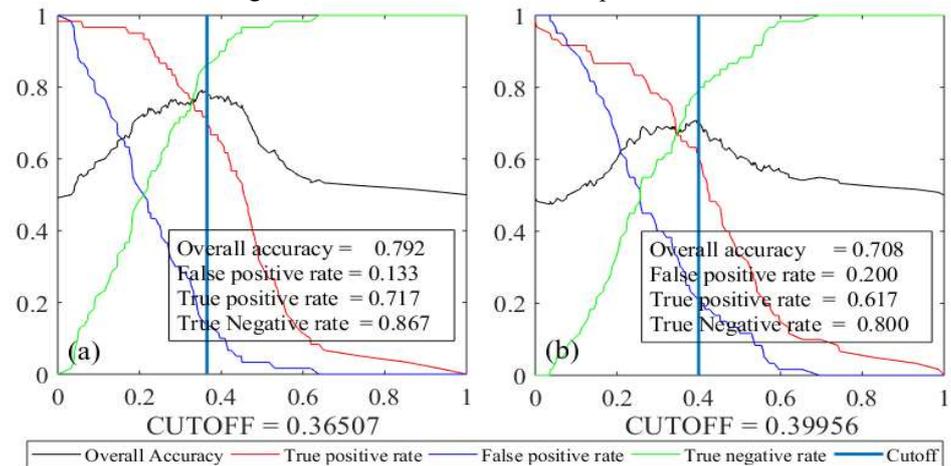

Fig. 11. Model-4 vehicle crash risk pre-detection

*2) Efficiency Measurement of Pre-detection*

To further understand the reliability of our proposed model, we apply the ROC curve indexes to evaluate the algorithm on drawing meaningful knowledge from the observed data and inferring the risk driving level for a given prediction time horizon. As shown in Figs. 8-11 that different overall accuracy, true positive rate, false positive rate and true negative rate are calculated by changing the cutoff value. In each figure, the vehicle crash risk predicted for specific prediction time from 0.5 second to 1 second are shown in left graph and right graph respectively.

In Fig. 8(a), 91.7% of risky driving behavior involved in near crash situation is correctly predicted 0.5s before the driver taking the harsh deceleration in those scenarios, while 96.7% of safety driving status are accurately identified, which means that the false warning rate is only 3.3%, the overall driving risk prediction accuracy is 94.2%. The risky driving behavior also has been predicted before 1s in Fig. 8(b), we target the 90.0% harsh deceleration behavior. It illustrates that prediction accuracy of Model-1 presents lightly reduction at longer time interval prediction.

We also examine the prediction accuracy of Model-2 and Model-3 in Fig. 9 and Fig. 10 respectively. The results show that the performance of Model-2 fluctuates when predicting the driving safety in longer time intervals. In Fig. 9(a), 93.3% of safety driving status are accurately identified, however, in Fig. 9(b), only 83.3% of true negative cases are effectively achieved, which means the increase of false warning for risk driving. In Fig. 10, although the overall accuracy of 86.7% and 85% respectively in 0.5s and 1s prediction time with very low false positive rate is considered reasonable, Model-3 performs not as good as the Model-1 and Model-2, the inclusion of abundant information may cause the overfitting in crash risk assessment. In Fig. 11, we achieved the lowest prediction accuracy by applying TTC model, only 61.7% of harsh braking and 80.0% of safety status is predicted before 1s. The results further indicate that the vehicle near-crash events can diversify into different driving safety level when having same headway (TTC) before driver making the effort, since the driver maneuver will influence the driving safety in most emergency cases. These results show that the VPRS model framework seems to be quite robust with respect to realistic vehicle near crash risk assessment.

## VI. CONCLUSIONS

In this paper, we proposed a data mining model based on systematic "driver-vehicle-road traffic" arrangement for evaluating the driving safety in near-crashes, which can provide effective decision and warning information for ADAS. The involvement of crash risk in certain emergency situation is linked with the real time conditional attributes (driver behavior, vehicle motion, etc.) through an improved rough set model, which can be trained and validated using driving data. The rules bases can then be used for evaluating crash risk tendency for a new case and setting conditional attributes to support risk assessment. A case study was conducted to identify the emergency situation in field test based on the proposed method. Our proposed methodology has the following features:

- The proposed method comprehensively analyzed the effective driver intention, vehicle motion and traffic environment on current driving safety, and quantified the vehicle crash risk (by predicting driver's real reflections in next step) in near-crash scenarios.
- The improved rough set model indicated the input-output relationships characterized by extracted rules, which are interpretable and easy to understand, while other models are black boxes in nature and their input-output relationships cannot be interpreted or verified explicitly. This transparent input-output relationships are very important for retro designing ADAS.
- The proposed method can further accommodate driver's experience. Although it has not been discussed in this paper. Expert knowledge can be incorporated in rough set based models as constraints, or the initial value of training parameters in the proposed model can be set by experts intuitively whenever possible, leading to an expert-data driven system.

It also should be noted that, there are some limitations in our conducted field driving test. In our current database, the influence of multi factors on the driving risk was not fully addressed. Only longitudinal driving safety situation assessment has been processed and evaluated. The time-duration of the current experiment was not very long enough to collect data under all conditions. Despite such limitations, the proposed VPRS quantify the driving risk in near-crash event and to analyze the associated risk-factors, this can be extrapolated to specific studies on other datasets.

As drivers with different personality may weigh safety, comfort, driving efficiency and other factors very differently, further research will consider the influence of driver's personality on driver behavior in near-crash situations. In this case, driver's personality (e.g., personal weights to above aspects) can be modeled by choosing different parameters for the related cost function, and the challenge is to design a suitable cost function reflecting different driving styles. Furthermore, other driving intension should be considered to capture more complex scenarios, such as lane change, overtaking and turn round etc. These scenarios can be constructed by the basic scenarios studied in this paper. Again, the main challenge is to design a suitable cost function that accurately represents corresponding driver behavior. Finally, it would be interesting to apply the proposed method for retro design of vehicle collision avoidance system to helps driver to take effective action before vehicle involves in high risk situation in real traffic near crash scenarios.

ACKNOWLEDGMENT

This work was jointly supported by National Key R&D Program of China (Grant No. 2017YFC0803900) and National Nature Science Foundation of China under the Grant No. 61703160, No. U1764262, No. 51775396 and No. 51605350. The authors would also like to thank those who participated in the driving experiments.

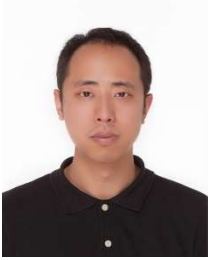

**Dr. Liqun Peng** is currently an associate professor with the School of Transport at the East China Jiaotong University. He received his Ph.D. degree in Transportation Engineering from Wuhan University of Technology, in China, in 2015. He has worked at University of Alberta as a Postdoctoral Fellow (2016-2017). His research interests are in the area of advanced driving assistance system, connected vehicle and traffic big data, specifically for improving roadway mobility and safety for both arterial and freeway. He has been awarded an Outstanding Ph.D. Thesis Award at the 10th Annual Meeting of China Intelligent Transportation Research Academic in 2015, and a Best Paper Award at the 13th Annual China Academic Conference for Ph.D. Candidates in 2015.

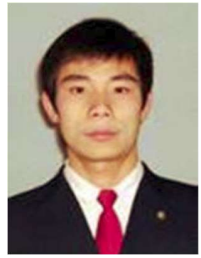

**Dr. Zhixiong Li** (M'16) received his Ph.D. degree in Transportation Engineering from Wuhan University of Technology, China. Currently he is a research associate in School of Mechanical, Materials, Mechatronic and Biomedical Engineering, University of Wollongong, Australia. His research interests include Intelligent Vehicles and Control, Loop Closure Detection, and Mechanical System Modeling and Control. He is an associate editor for the Journal of IEEE Access.

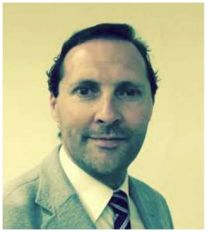

**Professor Miguel Angel Sotelo (Fellow IEEE)** received his Ph.D. degree in Electrical Engineering in 2001 from the University of Alcalá (UAH), Alcalá de Henares, Madrid, Spain. He is Head of the INVETT Research Group and Vice-President for International Relations at the University of Alcalá. He has been the Editor-in-Chief of IEEE Intelligent Transportation Systems Magazine, (2014-2016) and an Associate Editor of IEEE Transactions on Intelligent Transportation system (2008-2015). Currently, he is the President of the IEEE Intelligent Transportation Systems Society.

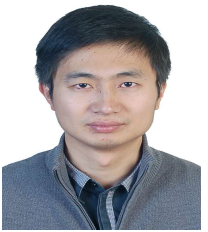

**Dr. Yi He** is currently a research fellow at California PATH, University of California, Berkeley, CA, USA. He received his Ph.D degree in Intelligent Transportation System Engineering from Wuhan University of Technology, China in 2015. His research interests include vehicle safety and driving behavior.

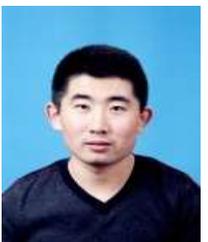

**Dr. Yunfei Ai** currently works as Postdoctoral Research Associate with the National Engineering Laboratory for Transportation Safety and Emergency Informatics in Beijing, China. He received the B.S. degree in Transport from Dalian Maritime University, Dalian, China, in 2011, the M.S. degree in Transportation Planning and Management from Dalian Maritime University, Dalian, China, in 2013, and the Ph.D. degree in Transportation Planning and Management from Dalian Maritime University, Dalian, China, in 2016. His research interests include transport safety, emergency management and Emergency Informatics.